\documentclass[12pt]{article}
\usepackage{algorithmic}
\usepackage{algorithm}
\usepackage{amsmath,amssymb,amsthm,mathrsfs,amsfonts}
\usepackage{bm}
\usepackage{latexsym}
\usepackage{CJK}
\usepackage{mathtools}
\usepackage{multirow}
\usepackage{multicol}
\usepackage{caption}
\usepackage{graphicx,color}
\usepackage{amssymb,amsfonts}
\usepackage{hyperref}
\usepackage{times}
\usepackage[T1]{fontenc}
\usepackage{enumerate}
\usepackage{geometry}
\usepackage{float}
\usepackage{appendix}
\usepackage{booktabs}
\usepackage{endnotes}
\usepackage{setspace}
\usepackage{stmaryrd}
\usepackage[numbers]{natbib}
\usepackage[english]{babel}
\usepackage{etoolbox}
\usepackage{adjustbox}
\allowdisplaybreaks

\usepackage[shortlabels]{enumitem}
\setlist[enumerate]{noitemsep}

\usepackage{amssymb,amsmath,amsthm,enumitem}
\usepackage[compact]{titlesec}
\titlespacing{\section}{0pt}{*1.5}{*0.5}
\titlespacing{\subsection}{0pt}{*1.5}{*0.5}
\titlespacing{\subsubsection}{0pt}{*1.5}{*0.5}

\defcitealias{leeseung}{Lee and Seung's (1999, 2001)} 

\AtBeginDocument{%
\setlength{\belowdisplayskip}{-2pt} \setlength{\belowdisplayshortskip}{-2pt}
\setlength{\abovedisplayskip}{-2pt} \setlength{\abovedisplayshortskip}{-2pt}
}

\begin{document}

\def\spacingset#1{\renewcommand{\baselinestretch}%
{#1}\small\normalsize} \spacingset{1}

\title{Dermoscopic Image Classification with Neural Style Transfer}

\medskip

\date{}

\author{Yutong Li $^{1}$ , Ruoqing Zhu $^{2}$, Annie Qu $^{3}$, Mike Yeh $^{4}$ \footnotemark[1]}
\bigskip

\maketitle

\begin{abstract}
Skin cancer, the most commonly found human malignancy, is primarily diagnosed visually via dermoscopic analysis, biopsy and histopathological examination. However, unlike other types of cancer, automated image classification of skin lesions is deemed more challenging due to the irregularity and variability in the lesions’ appearances. In this work, we propose an adaptation of the Neural Style Transfer (NST) as a novel image pre-processing step for skin lesion classification problems. We represent each dermoscopic image as a style image and transfer the style of the lesion onto a homogeneous content image. This transfers the main variability of each lesion onto the same localized region, which allows us to integrate the generated images together and extract latent, low-rank style features via tensor decomposition. We train and cross-validate our model on a dermoscopic data set collected and preprocessed from the International Skin Imaging Collaboration (ISIC) database. We show that the classification performance based on the extracted tensor features using the style-transferred images significantly outperforms that of the raw images by more than 10\%, and is also competitive with well-studied, pre-trained CNN models using transfer learning. Additionally, the tensor decomposition further identifies latent style clusters, which may provide clinical interpretation and insights.
\end{abstract}

\renewcommand{\thefootnote}{\fnsymbol{footnote}}

\footnotetext[1] {Yutong Li was with the Department of Statistics, University of Illinois at Urbana-Champaign, Champaign, IL 61820, USA. He is now with the Statistical Analytics Department, Novartis Pharmaceuticals, East Hanover, NJ 07040, USA (email: yutongli92@gmail.com); Ruoqing Zhu is with the Department of Statistics, University of Illinois at Urbana-Champaign, Champaign, IL 61820, USA (email: rqzhu@illinois.edu); Annie Qu is with the Department of Statistics, University of California, Irvine, Irvine, CA 92627, USA (email: aqu2@uci.edu); Mike Yeh is with Wolfram Research, Champaign, IL 61820, USA (stringtron@gmail.com).}


\newpage
\spacingset{1.45} 

\section{Introduction}
\label{sec:introduction}
Skin cancer is the most common type of cancer in the world, with millions of cases occurring globally each year \cite{schwartz2008skin}. Melanoma, the deadliest type of skin cancer, causes up to more than 10,000 deaths in the U.S. annually \cite{schwartz2008skin}. However, studies show that the early detection of melanoma is crucial and can significantly increase the survival rates of patients \cite{schwartz2008skin, madan2010non}. Dermoscopy is the examination of skin lesions with a dermoscope, which allows for enhanced inspection of skin lesions unobstructed by reflections of skin surfaces \cite{conforti2020dermo}. Studies have shown that this technology yields a higher diagnostic accuracy compared to standard photography \cite{yelamos2019dermo} when implemented by dermatologists. Nevertheless, the correct classification of lesion types highly depends on the clinical experiences and skills of each dermatologist, and is prone to inter-observer variation \cite{marks2000epidemiology}. Therefore, this has motivated many studies to develop more automated and accurate classification methods to supplement the dermatologists \cite{okur2018survey, ganster2001automated, barata2013two, codella2017deep, lopez2017skin}.

Automated skin lesion classification and machine learning research for skin cancer diagnostics have received much attention in the last decade, especially for image feature extraction \citep{nachbar1994abcd, ganster2001automated, barata2013two}. One major area of such work focuses on quantifying the ABCD rule (\textit{Asymmetry}, \textit{Border}, \textit{Color}, and \textit{Differential Structure}) \cite{ganster2001automated, kasmi2016abcd, nachbar1994abcd} or the 7-point checklist \cite{barata2013two, menzies2000surface}, which have been widely adopted by dermatologists and can serve as initial inputs for either a deterministic or model-based classification scheme. On the other hand, deep learning has found much success in image analysis applications with deep feature extraction capability such as the Convolutional Neural Network (CNN) \cite{ simonyan2014very, krizhevsky2012imagenet}, and has been effective in dermoscopic classification problems \cite{lopez2017skin, esteva2017dermatologist, litjens2017survey}. However, although CNN-based models provide powerful classification performance, the black-box nature of CNN models \cite{koh2017understanding, buhrmester2019analysis} makes them less interpretable from a clinical perspective.

Tensor decomposition has shown to be a great tool for image clustering and latent feature extraction due to its natural representation of images as higher order tensors \cite{kolda2009tensor}; for example, brain FRMI \cite{zhou2013tensor}, head and neck cancer \cite{lu2014spec}, and breast cancer \cite{tang2020tensor}. However, there has been a lack of literature using tensor decomposition for skin lesion classification problems. One reason behind this may be due to the heterogeneity of lesion shapes, locations, and sizes, which makes implementating tensor decomposition challenging since it is infeasible to stack the images together \cite{jukic2013tensor}. Therefore, this motivates us to consider a novel image registration method \cite{oliveira2014medical} to align the lesion information to effectively implement tensor decomposition. Medical image registration is very popular among radiology \cite{brant2012fundamentals} and brain FRMI \cite{zhou2013tensor}, primarily to register patient images across different time periods, angles, or modalities \cite{oliveira2014medical}. However, medical registration is still lacking for dermoscopic imaging \cite{navarro2018accurate}. In this paper, we propose modern techniques such as neural style transfer to bridge this gap.

Neural Style Transfer (NST) \cite{gatys2015neural} is a successful application using deep learning. The algorithm originally aims to extract the style of an art or image and transfer it onto another image (commonly referred to as the 'content' image) \cite{gatys2017controlling, jing2019neural}. Recently, NST \cite{gatys2015neural} has been integrated with the Generative Adverserial Network (GAN) \cite{yi2019gan} for data augmentation purposes \cite{shorten2019survey, mikolaj2018dataaug}, and has been used to generate artificial skin lesion images \cite{mikolaj2018dataaug, chi2018gan}. However, the fundamental purpose of these methods is not to process and register existing images, but rather focusing on creating new, synthetic images from existing images. Therefore, they do not address the issue of heterogeneity of lesion structures, as mentioned above. Additionally, these methods \cite{mikolaj2018dataaug, chi2018gan} do not enforce spatial control over the style-transfer process \citep{gatys2017controlling}, which may capture unwanted texture information, e.g., background skin texture. 

In this paper, we propose a novel adaptation of the NST algorithm \cite{gatys2015neural, gatys2017controlling} as an image pre-processing and registration step for dermoscopic images, and subsequently implement the CANDECOMP/PARAFAC (CP) decomposition \cite{kolda2009tensor} as a feature extraction method for dermoscopic images. We introduce a new paradigm to capture texture information of skin lesions for melanoma classification. In particular, we first transfer the style of a skin lesion to a homogeneous content image, constructed from the mean values of the red/green/blue (RGB) channels across all lesion images. We further enforce spatial control \cite{gatys2017controlling} over the transfer process with the implementation of segmentation masks. Thus, only the lesion information is transferred and any non-lesion related information is omitted during the process. Our proposed method ultimately extracts, magnifies, and normalizes all the lesions' positions in the same imaging region, which permits the successful implementation of higher-order tensor decomposition methods \cite{kolda2009tensor} to learn predictive latent style features successfully for more accurate classification performance.

Our contributions are mainly two-fold. First, we introduce a novel paradigm to extract and register lesion textures for classification purposes by implementing the style transfer framework that is mostly used in artwork \citep{gatys2015neural}. We further show that the style-transferred images yield image features that are significantly more predictive than the features obtained from raw images using tensor decomposition, and are competitive with pre-trained and fine-tuned end-to-end CNN models. Second, our proposed method provides additional representation of different lesion clusters according to their style/texture, which may benefit clinical interpretation and diagnostics. As opposed to the existing style-GAN \cite{yi2019gan} literature for data augmentation of skin lesion classification problems \cite{mikolaj2018dataaug, chi2018gan}, our proposed method does not perform data augmentation. Instead we develop a more interpretable formulation for end-to-end deep learning models while still achieving competitive classification performances. To the best of our knowledge, our work is the first study to extract and register the texture of skin lesions for general melanoma classification, and is the first work to apply tensor decomposition to lesion images for interpretable clustering and feature extraction.

\begin{figure}[h]
    \centering
    \includegraphics[scale=0.55]{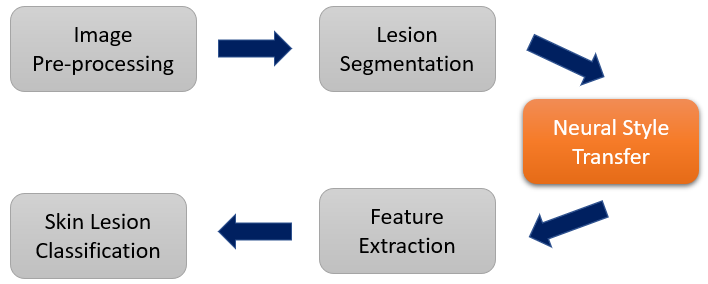}
    \caption{Analysis pipeline of this study.}
    \label{fig:pipe}
\end{figure}

The paper is structured as follow. Section \ref{sec:related} provides an overview of the existing literature. Section \ref{sec:dataset} presents a discussion of the dataset used in this study. Section \ref{sec:pre} discusses the pre-processing and segmentation of lesion images. Section \ref{sec:method} and \ref{sec:feature} present the proposed methodology and feature extraction methods respectively, along with the classification models considered for this study. We present the results of this study in Section \ref{sec:result}. Section \ref{sec:conclusion} concludes this study with some remarks on the proposed method. An illustration of our proposed analysis pipeline is provided in Figure \ref{fig:pipe}.

\subsection{Notations}

We first introduce the notation used in this study. Here, we denote a tensor as $\mathcal{X}$, where $\mathcal{X} \in \mathbb{R}^{I_1, I_2, ..., I_n}$, a matrix as $\textbf{X}$, and a vector as $\textbf{x}_i$. An image is denoted as $\Vec{x}$. In our study, each colorized dermoscopic image is a tensor with 3 color channels, and the corresponding lesion mask is a binary image with a single color channel. Additionally, $\odot$ is the element-wise product, $\ast$ is the Khatri-Rao product,  and $\dagger$ is the pseudoinverse. 

\section{Related Work}
\label{sec:related}
The existing literature in automated melanoma classification usually consists of three/four consecutive operations, i.e., (a) image pre-processing, (b) lesion segmentation, and (c) feature extraction and selection, and (d) classification \cite{okur2018survey, ganster2001automated}. 

Due to the innate limitations of dermoscopic imaging, unwanted artefacts such as dark corners, marker ink, gel bubbles, color discs, and ruler marks \cite{okur2018survey, conforti2020dermo} may be present. Another major artefact is the presence of skin hair \cite{conforti2020dermo}. Typically, median/bilinear filtering \cite{huang1979fast} has been widely adopted in removing minor image artefacts. For thicker skin hair, the DullRazor algorithm \cite{lee1997dullrazor} has been demonstrated to work well. On the other hand, different dermoscope devices and usage may cause variation in  illumination, contrast and noise \cite{ganster2001automated, barata2013two}. To normalize the color channels, one can subtract the average RGB value \cite{ganster2001automated} or intensity of the background skin \cite{codella2017deep} from each lesion. Alternatively, Barata \textit{et. al} \cite{barata2014color} investigated four different color normalization schemes \cite{barata2014color} for color and illumination constancy, and showed that these schemes can effectively improve classification performance.

The following step is to conduct image segmentation. This is done to isolate the lesion region from the background skin for downstream feature extraction. Although conceptually simple, this is a very important part of the automated classification pipeline, as most feature extraction methods (e.g. \textit{asymmetry} or \textit{border irregularity}), including the proposed NST algorithm, are dependent on the quality of the segmentation masks. The earliest and still widely used method for image segmentation is Otsu's method \cite{otsu1979mask}, which is simply performing image \textit{thresholding}; i.e., dividing the image into a foreground and background according to a specified pixel value. Later developments have include clustering (\textit{median cut}, \textit{k-means}, \textit{fuzzy-c means} and \textit{mean shift}) \cite{melli2006cluster} and more recently, deep learning based approaches \cite{yuan2017seg, codella2017deep, ronneberger2015u}. In particular, the fully connected U-net \cite{ronneberger2015u}, a symmetric U-shaped deep neural network, has proven to be very efficient and effective in performing medical image segmentation \cite{codella2017deep}. 

The next and most important step for automated lesion classification is feature construction and extraction, and consequently, classification \cite{ganster2001automated, barata2013two, codella2017deep, okur2018survey}. There are primarily two approaches for this step. The first approach is through manual feature construction, where different measurements of lesions are used as features, such as \textit{shape} (area, asymmetry, diameter, etc.), \textit{color} (mean and SD of color channels), and \textit{texture} (gray-level co-occurrence matrix \cite{pathak2013texture}). These quantifications are largely inspired by the ABCD rule \cite{nachbar1994abcd}, and have been widely adopted and expanded in the existing literature. For example, Ganster \textit{et al.} \cite{ganster2001automated} extracted color and shape features from images of a large dataset (5000+ images), and achieved a sensitivity and specificity of 87\% and 92\% respectively with a KNN classifier. Celebi \textit{et al.} \cite{celebi2007methodological} extracted a large feature vector for each lesion after dividing each lesion region into several clinical feature areas. They achieved a respective sensitivity and specificity of 93\% and 92\% on a unbalanced dataset of 564 images. Codella proposed a 2-systems approach, where they considered both global (extracted from entire lesion), and local (extracted from parts-of-whole lesion patches) features for classification on the $\text{PH}^2$ dataset of 200 images \cite{mendonca2015ph2}. Studies conducted by other researchers also follow this direction \cite{okur2018survey, zaq2019skin,  kasmi2016abcd, abbadi2017detection}. The second approach for feature extraction is to use CNN models \cite{lopez2017skin, esteva2017dermatologist, litjens2017survey, haenssle2018man}, usually under the transfer learning framework \cite{esteva2017dermatologist, pan2009survey}. Lopez \textit{et al.} considered transfer learning and fine tuning of the VGG16 network \cite{simonyan2014very} in the ISIC 2016 competition \cite{isic2016gutman}, and reported a new top sensitivity of 0.795. Esteva \textit{et al.} \cite{esteva2017dermatologist} presented a landmark work within this field, where they trained a large dataset of 129,450 images via transfer-learning using a pre-trained InceptionV3 model \cite{pan2009survey}. They reported an AUC of 0.94 for melanomas classified exclusively with dermatoscopic images. Codella \textit{et al.} \cite{codella2017deep} considered an ensemble of popular pre-trained models and hand-coded features, and reported a $7\%$ improvement over the top result in the ISIC 2016 competition \cite{isic2016gutman}. However, despite the promising results of these works, the different datasets used in each study make it difficult to compare the methods or relevance of the features across different studies \cite{codella2017isic, tschandl2018ham10000}. Recently, the International Skin Imaging Collaboration (ISIC) \cite{codella2017isic} has made an effort to unify the datasets used in dermoscopic imaging studies by providing a large repository of public images, and an annual data competition that consists of lesion segmentation, feature extraction, and classification \cite{isic2016gutman, isic2017codella}.

\section{Dataset}
\label{sec:dataset}

The dermoscopic images considered in this paper are described in \cite{tschandl2018ham10000}, and are available for public download through the ISIC database \cite{codella2017isic}. These are 24-bit JPEG images with a typical resolution of 768 × 512 pixels. However, not all the images in the database are in satisfactory condition. To first validate our approach, we constructed a high-quality, balanced dataset of 1000 images (500 malignant and 500 benign) by omitting the images that satisfiy any of the following conditions \cite{ganster2001automated}: (a) the entirety of the tumor does not fit within the image frame, or the tumor has been partially masked by the borders of the dermoscope, (b) an abundance of hair which blocks a significant portion of the lesion, to the extent that hair removal algorithms are ineffective, (c) are duplicated or augmented versions of other images. This data cleaning is necessary in order to ensure accurate border detection, reliable feature extraction, fair comparison of classification performances, and a satisfactory quality control for the style-transferred images without the interference of non-lesion information.

We then apply our proposed methodology to the lesion classification problems in both the ISIC 2016 \cite{isic2016gutman} and ISIC 2017 \cite{isic2017codella} competitions. The 2016 dataset contains 900 training images and 379 testing images with 2 lesion types, i.e., benign versus malignant \citep{isic2016gutman}. The 2017 dataset contains 2000 training images and 600 testing images with 3 lesion types, i.e., melanoma, nevus, and seborrheic keratosis \citep{isic2017codella}. The 2017 dataset is split into 2 problems: (a) melanoma (malignant) versus nevus and seborrheic keratosis together (benign), and (b) seborrheic keratosis (non-melanocytic) versus melanoma and nevus together (melanocytic). These results are presented in Section \ref{sec:result}.

\section{Image Pre-processing and Segmentation}
\label{sec:pre}
In this section, we first discuss the standard pre-processing of the lesion images, followed by image segmentation of the lesions to extract the corresponding segmentation masks. Next, we present a review of the NST method \cite{gatys2015neural}, followed by the proposed methodology to transfer the style of the lesions onto a homogenized content image using the guided-NST algorithm \cite{gatys2017controlling}. 

\subsection{Image Pre-processing}
 We first implement bilinear interpolation \cite{li2001new} to resize all the images to $224\times224\times3$ (input size of the VGG network \cite{simonyan2014very}) to reduce the computational complexity of the style-transfer algorithm, feature extraction, and the final classification problem. Next, we remove the air bubbles, thin hairs and other minor image artefacts using median filtering \cite{huang1979fast}. From our numerical experiments, we found that a $5\times5$ filtering window is a good choice for removing most image noises while still preserving the quality and resolution of the lesion. We then implement the DullRazor software \cite{lee1997dullrazor} to remove larger hair threads that were irremovable from the previous step.
 
\begin{figure}[h]
     \centering
     \includegraphics[scale=0.55]{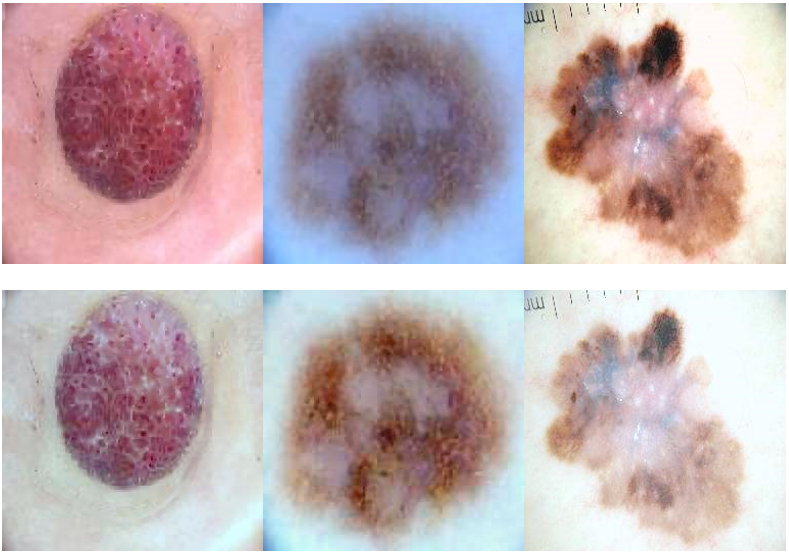}
     \caption{Top: Image before color normalization. Bottom: Image after color normalization via the Shades of Gray method \cite{finlayson2004gray}.}
     \label{fig:shadesofgray}
 \end{figure}

 Since the color of a lesion is dependent on the background skin color \cite{ganster2001automated}, and each dermoscopic image could be subjected to different device and illumination settings \cite{barata2014color}, it is necessary to normalize the color channels of each image. Here we implement the Shades of Gray algorithm \cite{finlayson2004gray} to normalize the images, as shown in Fig. (\ref{fig:shadesofgray}).
 
 \subsection{Image Segmentation}
 
 In order to both perform the guided-NST method \cite{gatys2017controlling} and to further extract skin lesion features, we need to first isolate the lesion from the skin information \cite{lopez2017skin, codella2017deep, litjens2017survey}. 
 
\begin{figure}[h]
     \centering
     \includegraphics[scale=0.7]{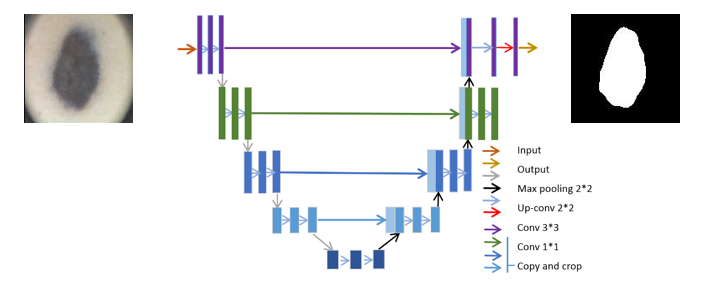}
     \caption{A U-net \cite{ronneberger2015u} architecture to learn and generate the segmentation mask.}
     \label{fig:unet}
 \end{figure}

 We perform image segmentation by implementing the U-net architecture \cite{ronneberger2015u}, as shown in Fig. \ref{fig:unet}, under the Tensorflow \cite{abadi2016tensorflow} framework. We use a convolution kernel size of $3\times3$ with batch normalization at the end of each of convolution block, max pooling layers of $2\times2$, 16 convolution filters at the first stage, a 256-dimensional fully connected layer, and a dropout of 0.5 at all dropout layers. We implement the Adam optimizer \cite{kingma2014adam} with an initial learning rate of 0.0005 and momentum of 0.95. We consider the binary cross-entropy loss for this U-net model.
 
 The U-net model is trained from the $\text{PH}^2$ dataset \cite{mendonca2015ph2}, with the dataset divided according to a $90\%/10\%$ training/validation split. We apply standard data augmentation procedures \cite{shorten2019survey} to increase the available training samples and to prevent overfitting, including size re-scaling, rotations of $40^{\circ}$, horizontal/vertical shift/flipping, image zooming, and shearing. The model is then trained on the training set for 200 epochs. Early stopping \cite{prechelt1998early} is implemented when the validation accuracy does not improve for at least 20 epochs to reduce the potential of overfitting. We then generate the segmentation mask for our dataset of 1000 images after the training is done.
 
\begin{figure}[h]
     \centering
     \includegraphics[scale=0.6]{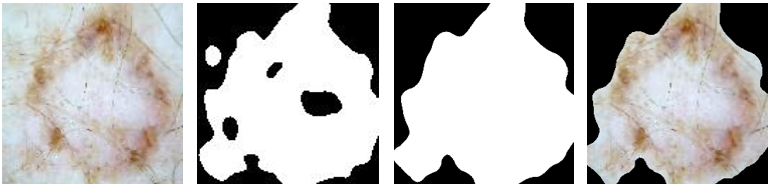}
     \caption{Post-processing of a generated lesion mask. The largest white blob is retained and all miscellaneous holes are filled in.}
     \label{fig:segm}
 \end{figure}

 The generated segmentation masks may contain multiple regions, holes, or rough edges. Additional post-processing is done to isolate the Region of Interest (ROI) \cite{goyal2018region} by retaining only the largest white color blob and filling in all holes on each of the segmented mask (Fig. \ref{fig:segm}). A 9x9 median-blur filter \cite{huang1979fast} is finally applied to the masks for edge smoothing. 
 
\section{Proposed Methodology}
\label{sec:method}

\subsection{Review of Neural Style Transfer} Motivated by the goal of texture synthesis \cite{jing2019neural}, the NST algorithm \cite{gatys2015neural} aims to learn the style of an image (typically an artwork or photo) and transfers it onto another piece of image. This newly rendered image generally contains the stylistic representation of the former image with the content of the latter image preserved. This is done by using the convolution layers of a pre-trained VGG19 network \cite{simonyan2014very} to extract feature maps at different convolution layers, where each feature map represents certain aspects of the original image.  

A stylized content image $\Vec{x}$ is generated by first initializing as a white noise image (or simply the content image itself), then simultaneously optimized with respect to both a content loss and style loss. Let $\Vec{p}$, $\Vec{a}$ and $\Vec{x}$ represent the content image, the style image and the target image, and $\textbf{P}^l$, $\textbf{A}^l$, $\textbf{F}^l$ be their respective feature representations in layer $l$. Here $\textbf{P}^l$, $\textbf{A}^l$, and $\textbf{F}^l$ all have dimensions $M_l \times N_l$, where $N_l$ is the number of feature maps in layer $l$ and $M_l = h_{l} \times w_{l}$ is the product of the height and width of each feature map in layer $l$. 

The content loss is defined as the squared loss between the feature representation of the content image and the target image, as given in (\ref{eq:cont}),
\begin{equation}
    \mathcal{L}_{\text{content}}(\Vec{p}, \Vec{x}, l) = \frac{1}{2}\sum_{i,j} (\textbf{F}^l_{ij} - \textbf{P}^l_{ij})^2.
    \label{eq:cont}
\end{equation}

Gatys \cite{gatys2015neural} defined the style representation as a correlation measure between the vectorized feature maps $\textbf{F}^l$ in layer $l$. These correlations are given by the Gram matrix $\textbf{G}^l \in \mathcal{R}^{N_l \times N_l}$, where $\textbf{G}^l_{ij}$ is the inner product between the vectorized feature map $i$ and $j$ in layer $l$, shown in (\ref{eq:gram}),
\begin{equation}
    \textbf{G}^l_{ij} = \sum_k \textbf{F}^l_{ik} \textbf{F}^l_{jk}.
    \label{eq:gram}
\end{equation}

\noindent The Gram matrix effectively utilizes the second-moment information of the feature maps in each considered convolution layer. As the feature maps are vectorized during the calculation, the exact spatial structures of the pixels are discarded and only the correlations between the feature maps are captured, which is desired for style transfer purposes as we do not want the exact "content" of the style image to be carried over. Moreover, Li et. al \cite{li2017demystifying} further showed that style of an image can be intrinsically represented by feature distributions in different layers of a CNN, and that style transfer can be seen as a distribution alignment process from the content image to the style image. The style loss is the squared loss between the style representation of the original style image $\Vec{a}$ and the target image $\Vec{x}$ in layer $l$. For each layer, the contribution to the total style loss $\mathcal{L}_{\text{style}}$ is given in (\ref{eq:styloss}),

\begin{equation}
    E_l = \frac{1}{4N^2_l M^2_l} \sum_{i,j} (\textbf{G}^l_{ij} - \textbf{A}^l_{ij})^2, 
    \label{eq:styloss}
\end{equation}
and the total style loss $\mathcal{L}_{\text{style}}$ is given in Equation (\ref{eq:totstyloss}),
\begin{equation}
    \mathcal{L}_{style}(\Vec{a},\Vec{x}) = \sum^L_{i=0} w_l E_l
    \label{eq:totstyloss}
\end{equation}
where $w_l$ are the weight factors of the contribution of each layer to the total loss. Finally, the extent to which the generated image resembles either the style or content images is controlled by two weight parameters $\alpha$ and $\beta$, as shown in (\ref{eq:genloss}),
\begin{equation}
    \mathcal{L}_{\text{total}} = \alpha \mathcal{L}_{\text{content}} + \beta\mathcal{L}_{\text{style}}.
    \label{eq:genloss}
\end{equation}

\subsection{Proposed Guided Style Transfer for Skin Lesions}
\begin{figure}[h]
    \centering
    \includegraphics[scale=0.7]{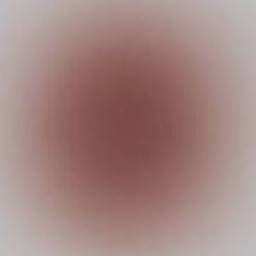}
    \caption{Content image generated by taking the average of the RGB channels across all lesion images in the data set.}
    \label{fig:content}
\end{figure}

We first regard each of the pre-processed lesion images as the base style images. For the content image, we took the average of each of the RGB channels across all the original lesion images (after median filtering and hair removal) in the data set to create a single content image (Fig. \ref{fig:content}). There are two particular reasons for the choice of this content image. First, this eliminates the necessity of the content image being an additional tuning parameter. Instead, the content image is deterministic, data-driven, and can be readily extended to the implementation of other lesion datasets. Second, our experiments shows that the style transfer process naturally transfers the main parts of the lesion information onto the target area without any tuning required.

To transfer the style of the lesion without the background skin information, we apply the guided-NST algorithm \cite{gatys2017controlling} under the Tensorflow framework \cite{abadi2016tensorflow} to transfer the style of each lesion onto the homogeneous content image. Here we utilize the segmentation masks we generated in the previous step to guide the style transfer process by masking the feature maps at each considered convolution layer of the style transfer process. To propagate the mask to the appropriate dimension at each considered layer, we construct a similar network structure as the VGG19 network \cite{simonyan2014very}, but without the convolution and ReLU layers. We only retain the pooling layers here as our goal is to perform dimension reduction of the segmentation mask. We denote the segmentation mask at layer $l$ of the VGG19 network as $\textbf{T}^l$, where $\textbf{T}^l \in \mathcal{R}^{h_l \times w_l}$. Each channel mask is normalized such that $\sum_i (\textbf{{T}}^l_i)^2 = 1$.

After the segmentation mask is generated for each convolution block, we then perform a Hadamard product between the columns of $\textbf{F}^l$, i.e., $\textbf{f}^l_j$ where $j \in (1,...,N_l)$ and the vectorized channel mask $\textbf{t}^l \in \mathcal{R}^{M_l \times 1}$ in one convolution layer, yielding the masked feature maps for that particular layer: 
\begin{equation}
    \Tilde{\textbf{f}}^l_{[j]} = \textbf{f}^l_{[j]} \odot \textbf{t}^l.
    \label{eq:maskedfeature}
\end{equation}
\noindent The new masked Gram matrix $\Tilde{\textbf{G}}^l$ is simply the dot product of the masked feature maps $\Tilde{\textbf{F}}^l$,
\begin{equation}
    \Tilde{\textbf{G}}^l_{ij} = \sum_k \Tilde{\textbf{F}}^l_{ik} \Tilde{\textbf{F}}^l_{jk}.
    \label{eq:maskedgram}
\end{equation}

The generated images may have undesirable artefacts and high frequency noise, especially for bright or dark pixels. Thus, in addition to the style loss and content loss, we also consider the total variation loss \cite{chan2005recent} for image smoothing. Assuming that $x_{i,j}$ represents the pixel value at coordinate $(i,j)$ of the target image, the total variation loss, denoted as $\mathcal{L}_{\text{tv}}$, is given as (\ref{eq:tvloss}).
\begin{equation}
    \mathcal{L}_{\text{tv}} = \sum_{i,j} |x_{i,j} - x_{i+1,j}| + |x_{i,j} - x_{i,j+1}|.
    \label{eq:tvloss}
\end{equation}
Thus, the final loss function to generate an image is
\begin{equation}
    \mathcal{L}_{\text{total}} = \alpha \mathcal{L}_{\text{content}} + \beta\mathcal{L}_{\text{style}} + \gamma \mathcal{L}_{\text{tv}},
    \label{eq:finaloss}
\end{equation}
where $\gamma$ is the weight of the total variation loss.

We perform a grid search across different style layers, content layers, style/content ratios, and total variation weights as shown in Table \ref{tab:grid} to examine the quality of the style transfer and how it impacts downstream classification performances. Both the style and content layers are chosen according to Gaty's work \cite{gatys2015neural}, except that we replace the convolution layers with the ReLU layers as they generate images with slightly better quality according to our experiments. 
\begin{table}[h]
    \centering
    \begin{tabular}{|l|l|}
    \hline
    Components             & Tuning Parameters                      \\ \hline
    Style Layers           & \begin{tabular}[c]{@{}l@{}} {[}relu1\_1{]},  {[}relu1\_1, relu2\_1{]},\\{[}relu1\_1, relu2\_1, relu3\_1{]},\\ {[}relu1\_1, relu2\_1, relu3\_1, relu4\_1{]},\\ {[}relu1\_1, relu2\_1, relu3\_1, relu4\_1, relu5\_1{]}\end{tabular} \\ \hline
    Content Layers         & {[}relu1\_2{]}, {[}relu2\_2{]}, {[}relu3\_2{]}, {[}relu4\_2{]}, {[}relu5\_2{]}  \\ \hline
    Style/Content Ratio    & 1, 10, 100, 1000, 10000          \\ \hline
    TV Weight & 1, 10, 100      \\ \hline
    \end{tabular}
    \caption{Grid search of tuning parameters for guided neural style transfer of skin lesions.}
    \label{tab:grid}
\end{table}
    
\noindent Due to the intense computation cost of the image generation process across all considered tuning parameters, the image generating algorithm is run on three separate GPUs, i.e., Nvidia K40, Nvidia RTX 2060 Super, and Nvidia Tesla V100, where each style transfer process is run for 500 iterations or if the relative change in loss is less than $0.0005$ given in (\ref{eq:reloss}),  

\begin{equation}
    \text{Relative Loss} = \frac{|\epsilon_t - \epsilon_{t-1}|}{\epsilon_t} \leq 0.0005,
    \label{eq:reloss}
\end{equation}
where $\epsilon_t$ is the loss at the $t$-th iteration. The GPUs do not affect the quality of the style transfer, only the speed of the generation process. Each image takes about 15 seconds, 9 seconds, and 5 seconds to generate respectively, with V100 being the fastest.

\begin{figure}[h]
    \centering
    \includegraphics[scale=0.7]{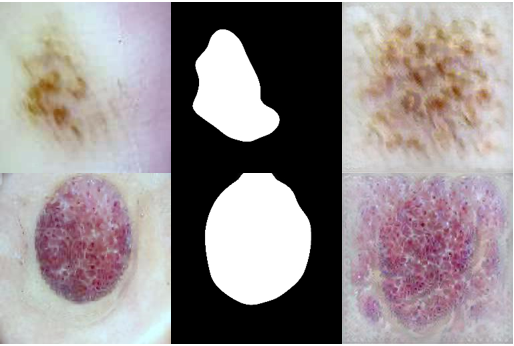} \\
    \vspace{2mm}
    \includegraphics[scale=0.7]{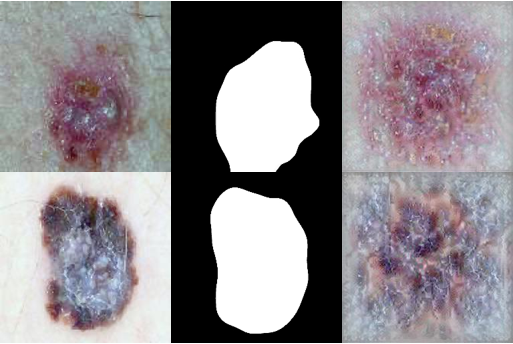}
    \caption{Examples of style-transferred image with spatial control using style masks. The top 2 are benign lesions and the bottom 2 are melanomas.}
    \label{fig:exam}
\end{figure}

We present 2 pairs of the style transferred images with style masks guidance in Fig. \ref{fig:exam} as examples, i.e., two for benign lesions and two for melanomas. The images shown here are based on the tuning parameters that yielded the best classification performance, which are reported in Section \ref{sec:result}. Additional selected images to illustrate the effects of different tuning parameters are provided in the Supplementary Materials.

From the images presented in Fig. \ref{fig:exam}, we observe that the guided-style transfer achieves two main objectives. First, the overall textures of the lesions are successfully transferred over to the content image. For example, the red/pinkish base, pale blisters, and minor brown spot on the third image are carried over to the content image. Similarly, the large batch of white speckles in the fourth image is widely preserved onto the content image, although the exact location of the white batch is scattered mostly to the corners of the image. This behavior is expected, as the Gram matrix does not account for the exact spatial locations of the textural features \cite{gatys2015neural}. Second, we observe that the style of each lesion is generally well-distributed across the content image regardless of its original shape. In particular, the brown streaks of the first image are well-preserved and scattered uniformly across the content image. This achieves the goal of normalizing the position of the texture of each lesion, which further permits the implementation of tensor decomposition, as discussed later in Section \ref{sec:feature}. 

\section{Feature Extraction and Classification Models}
\label{sec:feature}

\subsection{Feature Extraction}
In this subsection, we present the ABCD rule \cite{nachbar1994abcd,kasmi2016abcd}, followed by the tensor decomposition \cite{kolda2009tensor} for feature extraction. \\

\subsubsection{Asymmetry} 
We consider both the shape asymmetry and lengthening index \cite{nachbar1994abcd, kasmi2016abcd} to measure the degree of asymmetry of a lesion. \\

\noindent \textbf{Shape Asymmetry Index (SAI)}:  The ROI of each image obtained from the image segmentation step is first centered. Then, it is rotated by $\theta^{\circ}$ to align the $(x,y)$ coordinates with the principle axis of its centroid, where $\theta^\circ$ is defined as the angle between the x-axis and the axis around which the object can be rotated with minimum inertia, 
\begin{equation}
    \theta =\frac{1}{2} \arctan (\frac{2m_{11}}{m_{20} - m_{02}}).
    \label{eq:asym}
\end{equation}
\noindent The $m_{11}$, $m_{20}$, and $m_{02}$ here are the moments of inertia defined as $m_{pq} = \sum_i (x_i - x_0)^p(y_i - y_0)^q$, where $(x_0, y_0)$ is the centroid. An example of this process is provided below in Fig. \ref{fig:asymm}.

\begin{figure}[h]
    \centering
    \includegraphics[scale=0.38]{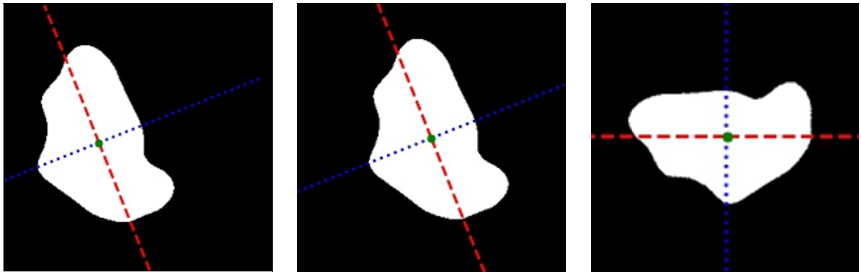}
    \caption{Example segmentation mask as shown in the first image of Fig. \ref{fig:exam}. From left to right: Original Mask, Centered Mask, Rotated Mask. The dot at the intersection of the two lines is the centroid of the ROI. The red dashed line is the major axis and the blue dotted line is the minor axis.}
    \label{fig:asymm}
\end{figure}

We then calculate the proportions of overlapping pixels of each ROI with respect to both principal axes. Let $A_{\text{top}}$, $A_{\text{bottom}}$ denote the top and bottom portion of the ROI with respect to the x-principal axis, and $A_{\text{left}}$, $A_{\text{right}}$ denote the left and right portion of the ROI with respect to the y-principal axis. We further denote $A^{x}_{\text{top}}$ as $A_{\text{top}}$ flipped with respect to the x-principal axis, and $A^{y}_{\text{left}}$ as $A_{\text{left}}$ flipped with respect to the y-principal axis. The SAI in terms of the x-principal axis and y-principal axis are given in (\ref{eq:sai}), 

\begin{equation}
    \text{SAI}_{x} = \frac{|A^{x}_{\text{top}} \cap A_{\text{bottom}}|}{|A^{x}_{\text{top}} \cup A_{\text{bottom}}|}, \text{ and } \text{SAI}_{y} = \frac{|A^{y}_{\text{left}} \cap A_{\text{right}}|}{|A^{y}_{\text{left}} \cup A_{\text{right}}|}.
    \label{eq:sai}
\end{equation}

\noindent \textbf{Lengthening Index}: This measurement describes the lengthening and the anisotropy degree of the skin lesion \cite{celebi2007methodological}, and is measured as the ratio between the eigenvalues $\lambda^{'}$ and $\lambda^{''}$ of the inertia matrix, as shown in \ref{eq:length},

\begin{equation}
    A = \frac{\lambda^{'}}{\lambda^{''}}, \text{ where} \\
    \label{eq:length}
\end{equation}
$$\lambda^{'} = \frac{m_{20} - m_{02} - \sqrt{(m_{20} - m_{02}) + 4(m_{11})^2}}{2}, $$
$$\lambda^{''} = \frac{m_{20} + m_{02} + \sqrt{(m_{20} - m_{02})^2 + 4(m_{11})^2}}{2}.$$ 

\subsubsection{Border Irregularity}
We compute the Compact Index, given in (\ref{eq:bi}) to measure border irregularity, defined as the ratio between the perimeter ($P$) and the area ($A$),
    \begin{equation} 
        \textcolor{black}{\textbf{BI} = \frac{P^2}{4 \pi A}}.
        \label{eq:bi}
    \end{equation}
\noindent The area ($A$)is the sum of all white pixels within each segmentation mask. For the perimeter ($P$), we slide a $3\times3$ convolution window of all 1s with only the white pixels as the center. The pixels that do not have a sum of 9, \textit{which implies that there exists at least one black pixel in its $3\times3$ neighborhood}, is considered as a border pixel. The perimeter ($P$) is then the sum of all such border pixels.\\
 
\subsubsection{Color}
\begin{figure}[h]
        \centering
        \includegraphics[scale=0.9]{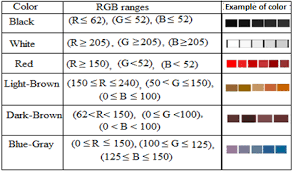}
        \caption{RGB thresholds for defining the color membership of each lesion \cite{abbadi2017detection}.}
        \label{fig:color}
    \end{figure}

Six different colors (white, red, light brown, dark brown, blue gray, and black) are considered according to the $\text{PH}^2$ database \cite{mendonca2015ph2, abbadi2017detection}. Thresholds for the pixel ranges of the red, green and blue channels constituting these six colors are given in Fig. \ref{fig:color}. We then compute the proportion of each color for each lesion as given from the table. 
     Additionally, we also compute the 7 summary statistics (\textit{min/Q1/median/Q3/max/mean/sd}) of the RGB channel of the images. \\

\subsubsection{Diameter} We construct a minimum bounding box over each rotated lesion mask as shown in Fig. 8. The pixel height and width of the bounding box is then used as the horizontal and vertical length of the lesion. \\

\subsubsection{Latent Style Features via Tensor Decomposition}
Now that the lesion styles are aligned on the same content image, we are able to stack all the images in the training set together and implement the CP decomposition \cite{kolda2009tensor} to extract latent style features across all training images. This step is to extract latent style/textures features for classification across all images in the training set. We first stack the color channels of each image onto one dimension ($224 \times 224 \times 3 \longrightarrow 224 \times 672$) to capture the interaction across different color channels. This allows us to perform a third-order tensor decomposition with respect to all the images in the training set. 

We apply the CP decomposition \cite{kolda2009tensor} on the training set $\mathcal{X} \in \mathcal{R}^{N_{train} \times 224 \times 672}$ to capture the global interaction between all the training images, where $N_{train}$ is the number of samples in the training set. This yields three low-rank factor matrices, $\textbf{A}_{train}$, $\textbf{B}_{train}$, and $\textbf{C}_{train}$ respectively. We then take out the image-factor matrix $\textbf{A}_{train} \in \mathcal{R}^{N_{train} \times R_{cp}}$ as the image features. Here, $R_{\text{cp}}$ is the specified rank of the decomposition, where we consider $R_{cp} = 24, 48, 72, 96$ for this study. After decomposing the training set, we then project the testing set onto the same column space as the training set to yield $\textbf{A}_{test}$, as shown in (\ref{eq:cp}). Let $\textbf{F} = \textbf{C}_{train} \ast \textbf{B}_{train}$, then

\begin{equation}
    \textbf{A}_{test} = \mathcal{X}_{(1)} \textbf{F} (\textbf{F}^{T} \textbf{F})^\dagger,
    \label{eq:cp}
\end{equation}

\noindent where $\mathcal{X}_{(1)}$ is the mode-1 unfolding of $\mathcal{X}$. We then use $\textbf{A}_{train}$ and $\textbf{A}_{test}$ as our features for classification.

\subsection{Classification Models}
\label{sec:class}
Here we discuss the supervised learning models considered in this study. For the image features constructed in Section 5.4, we consider the penalized linear regression \cite{glmnet2010}, random forest (RF) \cite{breiman2001random} and support vector machine (SVM) \cite{suykens1999least} for binary classification. We further consider different tuning parameters for each model in Table \ref{tab:model}, and report the best results out of all the supervised learning models and tuning parameters. 
\begin{table}[h]
    \centering
    \begin{tabular}{|l|l|}
    \hline
    \textbf{Models}                      & \textbf{Tuning Parameters}                                                                                                                                     \\ \hline
    Linear & $\alpha = 0, 0.5 ,1$                                                                                                                                    \\ \hline
    RF              & \begin{tabular}[c]{@{}l@{}}ntree = 500 (default)\\ mtry = 5, 10, $\sqrt{p}$, $p/3$ ($p =$ no. of features)\\ nodesize = 5, 10, 20, 30\end{tabular} \\ \hline
    SVM                         & \begin{tabular}[c]{@{}l@{}}kernel = linear, radial\\ cost = 0.01, 0.1, 1, 10\\ gamma = 0.01, 0.1, 1, 10 (radial kernel only)\end{tabular}               \\ \hline
    \end{tabular}
    \caption{Classic supervised learning models and considered tuning parameters.}
    \label{tab:model}
\end{table} 
We compare the results of our guided style transferred image with that of the raw images. Note that since the shape information is adversely discarded due to the style transfer process, all the ABCD features for the style-transferred images are computed from the original images to compensate for this loss in information. 

To show our method's competitiveness with end-to-end deep learning models, we also compare our performance via transfer learning \cite{pan2009survey} with two variants of the VGG network \cite{simonyan2014very}, i.e., the pre-trained VGG16 and VGG19 respectively. These two models are both trained on the pre-processed raw images. We follow the steps outlined in \cite{lopez2017skin}'s work to train both networks, i.e., (a) freezing all but the fully connected layers of the network, and (b) fine tuning the deeper parts of the network after Step (a). In addition, we also consider the same data augmentation procedures \cite{lopez2017skin, shorten2019survey} when training the U-net in the previous section.

We consider the accuracy, area under curve (AUC), sensitivity, and specificity \cite{powers2020eval} as metrics to evaluate classification performance. We perform a 5-fold cross-validation to evaluate each model, where the $80\%$ training and $20\%$ testing sets are split via random, stratified sampling. This cross-validation process is then further repeated 10 times with the results averaged across all runs.

\newpage

\section{Real Data Analysis and Results}
\label{sec:result}
In this section, we present the classification results of our analysis, followed by the interpretation of the style clusters according to the learned tensor features. 

\subsection{Classification Result}
We first present the results of the ABCD rule and tranfer learning of the raw images in Tables \ref{table:ABCD} and \ref{tab:vgg}, respectively. Next, we show the result comparison between the raw images and style-transferred images via (a) CP decomposition and (b) ABCD + CP decomposition in Table \ref{table:comp}. Note that the ABCD features for both the raw and generated images are derived from the raw images. 

\begin{table}[H]
\small
\centering
 \begin{tabular}{ llllll }
  \hline

    \multicolumn{3}{c}{\textbf{Raw Images (ABCD)}}\\
     
     \hline 
     \textit{Model} & \textit{Accuracy} & \textit{AUC} & \textit{Sensitivity} & \textit{Specificity} & \textit{Runtime} \\ 
      \hline
Linear & \begin{tabular}[c]{@{}l@{}}71.76\\ (3.21)\end{tabular} & \begin{tabular}[c]{@{}l@{}}79.14\\ (2.84) \end{tabular} & \begin{tabular}[c]{@{}l@{}}73.43\\ (3.66) \end{tabular} & \begin{tabular}[c]{@{}l@{}}70.10\\ (3.81) \end{tabular} & \begin{tabular}[c]{@{}l@{}}9.12\\ (1.84)\end{tabular} \\
RF     & \begin{tabular}[c]{@{}l@{}}71.51\\ (3.25)\end{tabular} & \begin{tabular}[c]{@{}l@{}}79.35\\ (3.18)\end{tabular} & \begin{tabular}[c]{@{}l@{}}72.43\\ (4.09)\end{tabular} & \begin{tabular}[c]{@{}l@{}}\textbf{70.60}\\ (4.03)\end{tabular} & \begin{tabular}[c]{@{}l@{}}0.97 \\ (0.11)\end{tabular} \\
SVM    & \begin{tabular}[c]{@{}l@{}}\textbf{72.05}\\ (3.17)\end{tabular} & \begin{tabular}[c]{@{}l@{}}\textbf{79.91}\\ (2.88)\end{tabular} & \begin{tabular}[c]{@{}l@{}}\textbf{76.38}\\ (4.09)\end{tabular} & \begin{tabular}[c]{@{}l@{}}67.73\\ (4.13)\end{tabular} & \begin{tabular}[c]{@{}l@{}}0.41\\ (0.16) \end{tabular} \\
     \hline
    \end{tabular}%
\caption{Mean and standard deviation of classification metrics and computation time for different classification models using the ABCD features for the \textbf{raw images}.}
\label{table:ABCD}
\end{table}

\begin{table}[H]
\small
\centering
 \begin{tabular}{ llllll }
  \hline

    \multicolumn{3}{c}{\textbf{Raw Images (Transfer Learning)}}\\
     
     \hline 
     \textit{Model} & \textit{Accuracy} & \textit{AUC} & \textit{Sensitivity} & \textit{Specificity} & \textit{Runtime} \\ 
      \hline
      \textbf{VGG16} \\
      \hline
\begin{tabular}[c]{@{}l@{}}FC Layers \\ Only\end{tabular}    & \begin{tabular}[c]{@{}l@{}}70.13\\ (3.06)\end{tabular} & \begin{tabular}[c]{@{}l@{}}76.84\\ (2.74)\end{tabular} & \begin{tabular}[c]{@{}l@{}}70.09\\ (3.97)\end{tabular} & \begin{tabular}[c]{@{}l@{}}70.33\\ (6.68)\end{tabular} & \begin{tabular}[c]{@{}l@{}}81.74\\ (26.67)\end{tabular}  \\ \hline
\begin{tabular}[c]{@{}l@{}}Fine Tune\\ Blocks 4\&5\end{tabular} & \begin{tabular}[c]{@{}l@{}}\textbf{73.33}\\ (2.64)\end{tabular} & \begin{tabular}[c]{@{}l@{}}\textbf{80.41}\\ (2.61)\end{tabular} & \begin{tabular}[c]{@{}l@{}}\textbf{72.46}\\ (3.77)\end{tabular} & \begin{tabular}[c]{@{}l@{}}\textbf{76.57}\\ (7.66)\end{tabular} & \begin{tabular}[c]{@{}l@{}}105.70\\ (23.09)\end{tabular} \\
       \hline
       \textbf{VGG19} \\
      \hline
\begin{tabular}[c]{@{}l@{}}FC Layers \\ Only \end{tabular}    & \begin{tabular}[c]{@{}l@{}}69.98\\ (3.52)\end{tabular} & \begin{tabular}[c]{@{}l@{}}75.38\\ (3.57)\end{tabular} & \begin{tabular}[c]{@{}l@{}}70.69\\ (4.08)\end{tabular} & \begin{tabular}[c]{@{}l@{}}68.97\\ (7.34)\end{tabular} & \begin{tabular}[c]{@{}l@{}}89.81\\ (39.29)\end{tabular}  \\ \hline
\begin{tabular}[c]{@{}l@{}}Fine Tune\\ Blocks 4\&5\end{tabular} & \begin{tabular}[c]{@{}l@{}}71.58\\ (2.79)\end{tabular} & \begin{tabular}[c]{@{}l@{}}78.87\\ (3.11)\end{tabular} & \begin{tabular}[c]{@{}l@{}}69.44\\ (3.14)\end{tabular} & \begin{tabular}[c]{@{}l@{}}74.67\\ (8.38)\end{tabular} & \begin{tabular}[c]{@{}l@{}}162.45\\ (71.89)\end{tabular} \\
       \hline
    \end{tabular}%
\caption{Mean and standard deviation of classification metrics and computation time for transfer learning with the VGG16 and VGG19 networks on the \textbf{raw images}. \textbf{FC Layers Only} freezes all weights except the FC layer. \textbf{Fine Tune Blocks 4 \& 5} unfreezes the 4th and 5th block in addition to the FC layer for further training and fine-tuning of the weights.}
\label{tab:vgg}
\end{table}

\begin{table}[H]
\small
\centering
 \begin{tabular}{ llllll }
  \hline

    \multicolumn{6}{c}{\textbf{Style-transferred Images vs. Raw Images}}\\
     
     \hline 
     \textit{Model} & \textit{Accuracy} & \textit{AUC} & \textit{Sensitivity} & \textit{Specificity} & \textit{Runtime} \\ 
      \hline
     \textbf{CP (Style)} \\
      \hline
    Linear & \begin{tabular}[c]{@{}l@{}}74.40\\ (2.92)\end{tabular} & \begin{tabular}[c]{@{}l@{}}81.54\\ (3.05) \end{tabular} & \begin{tabular}[c]{@{}l@{}}75.48\\ (3.40) \end{tabular} & \begin{tabular}[c]{@{}l@{}}73.32\\ (3.44) \end{tabular} & \begin{tabular}[c]{@{}l@{}}2.38\\ (0.43)\end{tabular} \\
    RF     & \begin{tabular}[c]{@{}l@{}}75.20\\ (3.25)\end{tabular} & \begin{tabular}[c]{@{}l@{}}81.31\\ (3.18)\end{tabular} & \begin{tabular}[c]{@{}l@{}}\textbf{77.85}\\ (4.09)\end{tabular} & \begin{tabular}[c]{@{}l@{}}72.55\\ (4.03)\end{tabular} & \begin{tabular}[c]{@{}l@{}}1.37\\ (0.11)\end{tabular} \\
    SVM    & \begin{tabular}[c]{@{}l@{}}\textbf{76.61}\\ (3.17)\end{tabular} & \begin{tabular}[c]{@{}l@{}}\textbf{82.72}\\ (2.88)\end{tabular} & \begin{tabular}[c]{@{}l@{}}77.10\\ (3.54)\end{tabular} & \begin{tabular}[c]{@{}l@{}}\textbf{75.05}\\ (4.13)\end{tabular} & \begin{tabular}[c]{@{}l@{}}0.72\\ (0.16) \end{tabular} \\
       \hline
       \textbf{ABCD+CP (Style)} \\
      \hline
    Linear & \begin{tabular}[c]{@{}l@{}}76.09\\ (3.18)\end{tabular} & \begin{tabular}[c]{@{}l@{}}83.20\\ (2.87) \end{tabular} & \begin{tabular}[c]{@{}l@{}}76.65\\ (3.62) \end{tabular} & \begin{tabular}[c]{@{}l@{}}75.52\\ (3.82) \end{tabular} & \begin{tabular}[c]{@{}l@{}}20.05\\ (2.65)\end{tabular} \\
    RF     & \begin{tabular}[c]{@{}l@{}}77.04\\ (2.93)\end{tabular} & \begin{tabular}[c]{@{}l@{}}83.22\\ (3.10)\end{tabular} & \begin{tabular}[c]{@{}l@{}}79.42\\ (3.99)\end{tabular} & \begin{tabular}[c]{@{}l@{}}74.65\\ (3.79)\end{tabular} & \begin{tabular}[c]{@{}l@{}}3.12\\ (0.28)\end{tabular} \\
    SVM    & \begin{tabular}[c]{@{}l@{}}\textbf{77.70}\\ (3.37)\end{tabular} & \begin{tabular}[c]{@{}l@{}}\textbf{84.13}\\ (2.94)\end{tabular} & \begin{tabular}[c]{@{}l@{}}\textbf{79.62}\\ (4.16)\end{tabular} & \begin{tabular}[c]{@{}l@{}}\textbf{75.78}\\ (3.89)\end{tabular} & \begin{tabular}[c]{@{}l@{}}0.69\\ (0.03) \end{tabular} \\
        \hline
       \textbf{CP (Raw)} \\
      \hline
Linear & \begin{tabular}[c]{@{}l@{}}64.08\\ (2.91)\end{tabular} & \begin{tabular}[c]{@{}l@{}}70.09\\ (3.42) \end{tabular} & \begin{tabular}[c]{@{}l@{}}63.88\\ (4.30) \end{tabular} & \begin{tabular}[c]{@{}l@{}}64.28\\ (3.00) \end{tabular} & \begin{tabular}[c]{@{}l@{}}2.99\\ (1.04)\end{tabular} \\
RF     & \begin{tabular}[c]{@{}l@{}}62.10\\ (2.67)\end{tabular} & \begin{tabular}[c]{@{}l@{}}65.74\\ (3.21)\end{tabular} & \begin{tabular}[c]{@{}l@{}} 58.10\\ (4.94)\end{tabular} & \begin{tabular}[c]{@{}l@{}}66.10\\ (4.32)\end{tabular} & \begin{tabular}[c]{@{}l@{}}2.88\\ (0.20)\end{tabular} \\
SVM    & \begin{tabular}[c]{@{}l@{}}65.06\\ (2.92)\end{tabular} & \begin{tabular}[c]{@{}l@{}}71.75\\ (2.82)\end{tabular} & \begin{tabular}[c]{@{}l@{}}66.80\\ (4.30)\end{tabular} & \begin{tabular}[c]{@{}l@{}}63.32\\ (3.87)\end{tabular} & \begin{tabular}[c]{@{}l@{}}0.51\\ (0.05) \end{tabular} \\
       \hline
       \textbf{ABCD+CP (Raw)} \\
      \hline
Linear & \begin{tabular}[c]{@{}l@{}}71.39\\ (2.57)\end{tabular} & \begin{tabular}[c]{@{}l@{}}79.13\\ (2.61) \end{tabular} & \begin{tabular}[c]{@{}l@{}}72.58\\ (3.34) \end{tabular} & \begin{tabular}[c]{@{}l@{}}70.20\\ (3.69) \end{tabular} & \begin{tabular}[c]{@{}l@{}}14.08\\ (2.81)\end{tabular} \\
RF     & \begin{tabular}[c]{@{}l@{}}71.38\\ (3.47)\end{tabular} & \begin{tabular}[c]{@{}l@{}}79.14\\ (2.98)\end{tabular} & \begin{tabular}[c]{@{}l@{}}70.20\\ (5.81)\end{tabular} & \begin{tabular}[c]{@{}l@{}}72.55\\ (4.08)\end{tabular} & \begin{tabular}[c]{@{}l@{}}1.50\\ (0.08)\end{tabular} \\
SVM    & \begin{tabular}[c]{@{}l@{}}71.94\\ (3.00)\end{tabular} & \begin{tabular}[c]{@{}l@{}}79.96\\ (2.72)\end{tabular} & \begin{tabular}[c]{@{}l@{}}74.72\\ (4.09)\end{tabular} & \begin{tabular}[c]{@{}l@{}}69.15\\ (3.74)\end{tabular} & \begin{tabular}[c]{@{}l@{}} 0.58\\ (0.04) \end{tabular} \\
        \hline
    \end{tabular}%
\caption{Mean and standard deviation of classification metrics and computation time between \textbf{style-transferred images} and \textbf{raw images} using 1) CP features and 2) CP + ABCD features.}
\label{table:comp}
\end{table}

Tables \ref{table:ABCD} - \ref{table:comp} indicate that the ABCD rule and transfer learning trained on the raw images provides $71.5\% - 73.3\%$ classification accuracy, with the fine-tuned VGG16 network performing the best. On the other hand, Table \ref{table:comp} shows that CP features derived from the generated images yield up to $76.76\%$ accuracy, which further yields $77.70\%$ by adding the ABCD features. Using the CP features extracted from the generated images alone outperforms the ABCD rule and VGG16 by $3\% - 5\%$. On the other hand, using the CP features derived from the raw images yield subpar results with only $65.06\%$. Furthermore, adding the CP features of the raw images to the ABCD features yields no improvement. Ultimately, we have gained more than 11\% improvement in classification performances when comparing the CP features of the generated images (76.61\%) to the raw images (65.06\%). This shows that stacking the lesion images together without any alignment will affect the quality of features learned by the CP decomposition. The above results support our hypothesis, in that registering the images to a homogeneous background yields significantly more predictive features via the CP decomposition.

We next present the results of our proposed method in both the ISIC 2016 \cite{isic2016gutman} and ISIC 2017 \cite{isic2017codella} competitions on the lesion classification tasks. We pre-processed the entire training and testing sets using the steps outlined in Section \ref{sec:pre}. For the style transfer process, we use the tuning parameters (Table \ref{tab:grid}) that yield the best result for our balanced dataset of 1000 images, i.e., with all 5 style layers, content layer `relu4\_2', style/content ratio 1000, and total variation weight 1. The AUC of our results are given below in Table \ref{tab:isic}.

\begin{table}[H]
\small
\centering
\begin{tabular}{llll}
\hline 
          & \textbf{ISIC 2016} & \textbf{ISIC 2017 (a)} & \textbf{ISIC 2017 (b)} \\ \hline
ABCD      & 78.82     & 72.30             & 80.36             \\
CP        & 77.47     & 76.89             & 85.90             \\
ABCD + CP & 79.70     & 78.36             & 88.83             \\
\hline
\end{tabular}
\caption{The AUC of the proposed pipeline on both the ISIC 2016 and ISIC 2017 competitions. Note that the ISIC 2017 competition has 2 problems: 1) Melanoma versus nevus + seborrheic keratosis and 2) Seborrheic keratosis versus melanoma + nevus.}
\label{tab:isic}
\end{table}
        
Our proposed method performed relatively well on both competitions, with the combination of both ABCD and CP features performing the best. According to the competition leaderboard, our result of 79.70\% on the ISIC 2016 dataset is comparable to the 2nd place of the 2016 competition \cite{isic2016gutman}. Likewise, our \textbf{averaged} AUC of 83.60\% across problem (a) and (b) of the ISIC 2017 competition is comparable to the 8th place on the leaderboard \cite{isic2017codella}. To our knowledge, the top performing models in the competition typically uses transfer learning and data augmentation, some even with additional external data sources during training \cite{isic2017codella}. Thus, the above results further show the competitiveness of our proposed method with the most popular existing frameworks, despite our approach being fundamentally different and without relying on any external data augmentation methods. 

\subsection{Contribution of Style Layers}

\begin{figure}[H]
    \centering
    \includegraphics[scale=0.6]{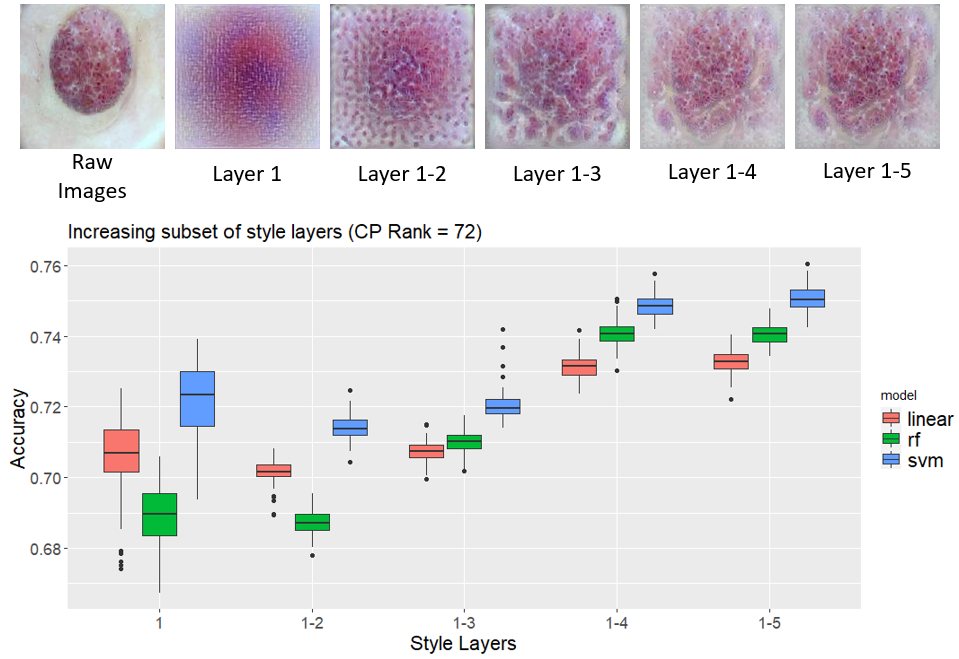}
    \caption{Generated images and classification accuracy box plots across different subsets of style layers.}
    \label{fig:bx_style}
\end{figure}

We next discuss the impact of different style layers on the classification performance of our balanced dataset. Recall the tuning parameters described in Table \ref{tab:grid}. In our experiments, we generate 375 sets of images based on the different combinations of tuning parameters and perform the same classification pipeline. To see how each tuning parameter contributes to the accuracy, we perform a regression with the 375 accuracy measures as the response and the different tuning parameters as categorical predictors. Our analysis shows that the most important tuning parameter is the style layer (p-value < $10^{-3}$), followed by the style-content weight ratio (p-value < $0.1$). The total variation weight and content layer are not significant. Here we present the box plot of the classification accuracy for the three models across the different subsets of style layers in Figure \ref{fig:bx_style}. 

\noindent The boxplots in Figure \ref{fig:bx_style} show that the classification accuracy improves as the number of considered style layers increases, with the highest accuracy using all the 5 style layers. The classification accuracy appears to be mainly driven from the first four layers. We observe that the texture of the lesion gets progressively more macroscopic and distinct as the number of style layers increases. For example, the main information in layer 1 consists of the color hue of the images but not the texture information. As more style layers are taken into account, the general texture starts to appear, starting from minor dots to speckles to more sophisticated patterns. This progressive change is the most evident for the first four layers, while adding the fifth layer does not provide a noticeable difference in the generated images, which explains the minor difference in improvement in classification accuracy as shown in Figure \ref{fig:bx_style}. Additionally, we notice that apart from style layer 1, the boxplots of the other style layers are quite flat. This implies that besides some statistical outliers, the other tuning parameters do not significantly impact the classification results. 

\subsection{Visualization of Tensor Clusters}
\begin{figure}[h]
    \centering
    \includegraphics[scale=0.46]{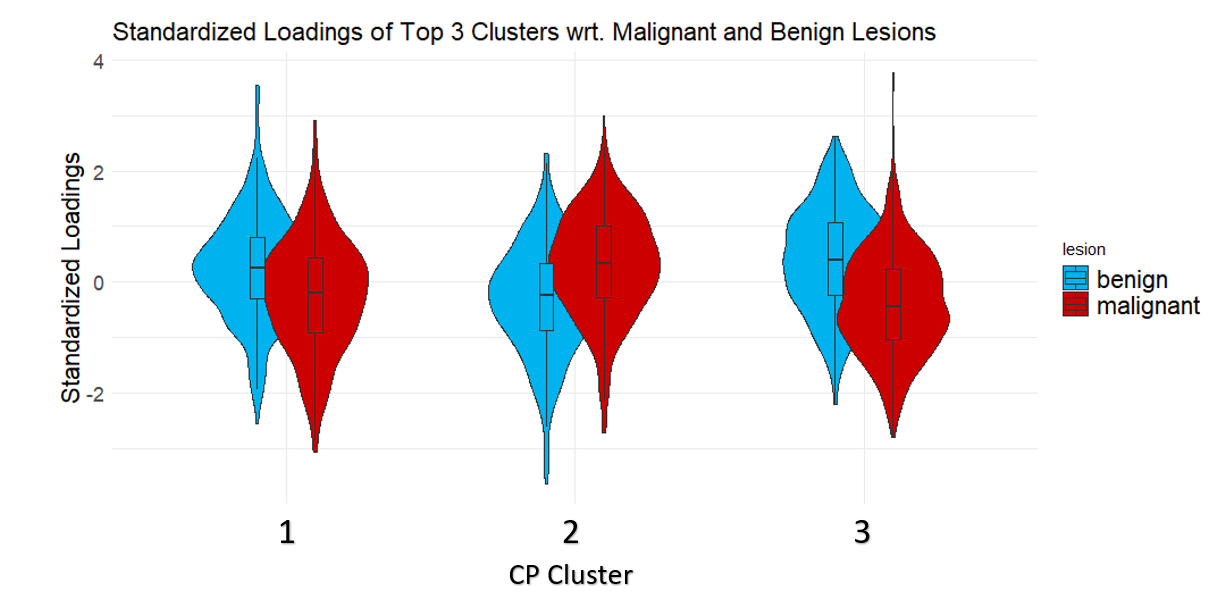}
    \caption{Standardized loadings of top 3 clusters as ranked by random forest with respect to malignant versus benign lesions. The blue and red violin plots represent the loadings of the benign and malignant lesion classes, respectively.}
    \label{fig:top3}
\end{figure}

We can further interpret the loadings of the $\textbf{A}$ matrix computed from the CP decomposition of the generated image to visualize the underlying style clusters. We illustrate this example in Figure \ref{fig:top3} by presenting violin plots to show the distribution of the standardized loadings for the top 3 clusters with respect to malignant versus benign lesions. These top 3 clusters are selected according to their variable importance from a random forest model with default settings.

\begin{figure}
    \centering
    \includegraphics[scale=0.5]{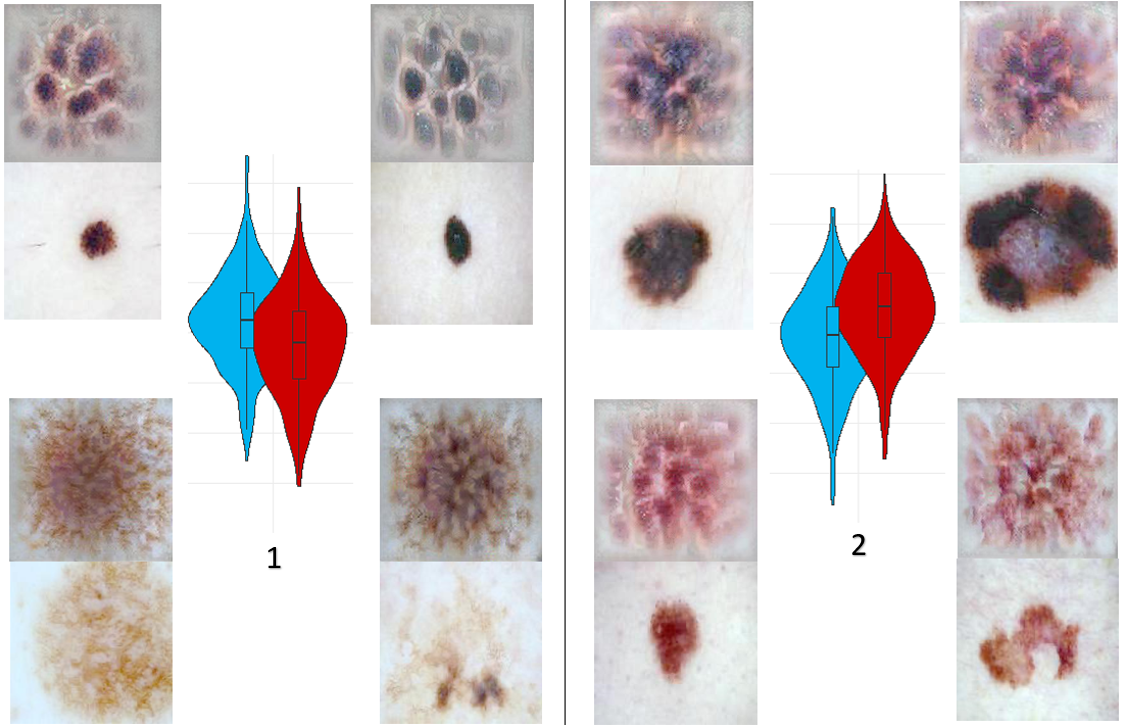}
    \caption{Representative generative images for the top 2 CP clusters. For each cluster, the two pairs of images at the top have the largest positive standardized loadings, whereas the two pairs of images at the bottom have the smallest negative standardized loadings. Within each image pair, the image at the top is the stylized image whereas the bottom is the original image.}
    \label{fig:interpret}
\end{figure}

From the violin plot, we observe that the normalized loading for the malignant tumors are visibly lower than the benign tumors for the first and third cluster, and higher than the latter in the second cluster. To interpret what each cluster represents, we examine the generated images with the largest positive loadings under the top two cluster vector in $\textbf{A}$, as shown in Figure \ref{fig:interpret}. These images can be regarded as the representative images of this cluster. Furthermore, we can also examine the images with the smallest negative loadings. These sets of images can be viewed as a contrast to the images with the positive loadings. To better understand these points, we present representative images in Figure \ref{fig:interpret} for the first two cluster vectors as selected by the random forest. For each cluster, four sets of images are presented, two for the largest postive loadings and two for the smallest negative loadings. The original source image for each generated image is also provided alongside.

From Figure \ref{fig:interpret}, we observe that the images with the largest positive loadings are scattered with small islands, whereas the images on the opposite spectrum are scattered with bouts of brown streaks and spots with a brown color center. From a visual standpoint, the CP decomposition adequately captures the characteristics of the generated images. Comparing with the corresponding original images, we notice the top loadings represent images with small sizes, smooth color textures, and have compact and uniform shapes. On the other hand, the images with negative loadings are large, have uneven color textures, and do not have a uniform boundary. Along with the standardized loadings, we can interpret that benign lesions typically have characteristics of the lesions with the largest positive loadings, whereas malignant lesions are more likely to have features similar to those of the smallest negative loadings with more uneven color texture. Likewise, the second cluster vector follows a similar interpretation. The above examples show that the CP decomposition on stylized images effectively provides an interpretable understanding of the different types of lesions from a style-driven perspective.

\section{Conclusion}
\label{sec:conclusion}
We proposed a novel method to register skin lesions onto a homogeneous background using neural style transfer to utilize tensor decomposition for better classification performances. Our method is conceptually simple and straightforward to implement, requiring only the lesion image and its corresponding mask, while at the same time yielding classification results as competitive as popular deep learning methods. Furthermore, the features learned from the tensor decomposition provide additional interpretability due to its innate cluster representation, which may be beneficial for clinical purposes. 

There are several potential directions for improvement. First, even though our method addresses the issue of different lesion shapes, locations, and sizes, our method also adversely forgoes this information during the transfer process. This may discard predictive information that can potentially improve classification performance. Therefore, it may be desirable to consider more flexible formulations, where weights may be assigned to the border pixels of feature maps during the calculation of the Gram matrix. Conceptually, this is similar to implementing soft masks on the lesion part, with different values on the border of the lesion as opposed to the core of the lesion. However, the choice of how to choose these mask values is nor trivial, and may require an elaborative optimization strategy to find the optimal weights. Another interesting direction is to consider utilizing the Gram matrix to directly extract texture features. This can be regarded as a new representation compared to the existing gray-level co-occcurence matrix \cite{pathak2013texture} for textural feature extraction, and may likely yield better results. Finally, there has been a lot of development of algorithmic advancements in neural style transfer \cite{jing2019neural}, and these advancements could be implemented for better image generation and classification results. In our case, we implement the guided NST \cite{gatys2017controlling} due to its flexibilty and established practicability.

Our work introduces a useful paradigm for registering and classifying dermoscopic images, which can serve as a stepping stone for future work on dermoscoping imaging and even other medical imaging data.

\end{document}